\documentclass{mn2e}
\usepackage{astrojournals}
\usepackage{epsfig,natbib2,natbibmnfix}
\usepackage{amssymb}
\usepackage{amsbsy}
\usepackage{multirow}
\usepackage{multicol}
\usepackage[varg]{txfonts}
\usepackage{color}
\usepackage{mathrsfs}
\usepackage{subfigure}

\newcommand{\cref}[1]{Chapter~\ref{#1}}
\newcommand{\sref}[1]{section~\ref{#1}}

\def \d{\mathrm{d}}
\def \b{\mathrm{b}}
\def \mathbi#1{\textbf{\em #1}}

\def\del#1{{}}

\begin{document}
\title[Stable counteralignment]{Stable counteralignment of a circumbinary
  disc} \author[C.J.~Nixon] {
\parbox{5in}{Christopher~J.~Nixon$^{1\star}$}
\vspace{0.1in} \\ $^1$ Department of Physics \& Astronomy, University of
Leicester, Leicester LE1 7RH UK} \maketitle

\begin{abstract}
In general, when gas accretes on to a supermassive black hole binary it is
likely to have no prior knowledge of the binary angular momentum. Therefore a
circumbinary disc forms with a random inclination angle $\theta$ to the
binary. It is known that for $\theta < 90^{\circ}$ the disc will coalign wrt
the binary. If $\theta > 90^{\circ}$ the disc wholly counteraligns if it
satisfies $\cos\theta < -J_{\rm d}/2J_{\rm b}$, where $J_{\rm d}$ and $J_{\rm
  b}$ are the magnitudes of the disc and binary angular momentum vectors
respectively. If however $\theta > 90^{\circ}$ and this criterion is not
satisfied the same disc may counteralign its inner regions and, on longer
timescales, coalign its outer regions. I show that for typical disc
parameters, describing an accretion event on to a supermassive black hole
binary, a misaligned circumbinary disc is likely to wholly co-- or
counter--align with the binary plane. This is because the binary angular
momentum dominates the disc angular momentum. However with extreme parameters
(binary mass ratio $M_2/M_1 \ll 1$ or binary eccentricity $e \sim 1$) the same
disc may simultaneously co-- {\it and} counter--align. It is known that
coplanar prograde circumbinary discs are stable. I show that coplanar
retrograde circumbinary discs are also stable. A chaotic accretion event on to
an SMBH binary will therefore result in a coplanar circumbinary disc that is
either prograde or retrograde with respect to the binary plane.
\end{abstract}


\footnotetext[1]{E-mail: chris.nixon@leicester.ac.uk}

\section{Introduction}
\label{intro}
Galaxy mergers are commonly thought to be the main mechanism driving the
coevolution of galaxies and their central supermassive black holes (SMBH). An
SMBH binary is likely to form in the centre of the merged galaxy and
subsequently accrete from a circumbinary disc. It is reasonable to expect that
the angular momentum of the binary and that of the accreting gas are
uncorrelated, and that each has no prior knowledge of the other. We can
therefore expect a random distribution of orientations for such circumbinary
discs.

Understanding the evolution of a misaligned circumbinary disc is needed if we
are to understand the evolution of SMBH binaries. In a recent paper
\citep{Nixonetal2011a} we showed that a retrograde circumbinary disc can be
very efficient in extracting angular momentum from the binary orbit. This may
offer a solution to the final parsec problem (\citealt{Begelmanetal1980};
\citealt{MM2001}).

In another recent paper \citep*[][hereafter NKP]{Nixonetal2011b} we showed
that the dominant effect of the binary potential, on a misaligned circumbinary
disc, is to induce radially--dependent precessions of the misaligned disc
particle orbits. This differential precession is known to induce warping of
the disc as rings of gas dissipate energy through viscosity. The precession
vanishes only when the orbit is in the plane of the binary (either prograde or
retrograde) and thus we expect that both prograde and retrograde orbits are
possible equilibria for the gas. NKP showed that the evolution of a misaligned
circumbinary disc is formally similar to the evolution of a misaligned disc
around a spinning compact object. In this case the evolution is driven by the
Lense--Thirring effect (e.g. \citealt{BP1975}; \citealt{Pringle1992};
\citealt{SF1996}; \citealt{LP2006}; \citealt{NK2012}). This implies that the
analysis of \citet{Kingetal2005}, which calculates the conditions for co-- or
counter--alignment, holds for circumbinary discs as well. The disc co-- or
counter--aligns depending on the magnitudes and directions of $\mathbi{J}_{\rm
  b}$ and $\mathbi{J}_{\rm d}$, the angular momentum of the binary and disc
respectively. The whole disc counteraligns with the binary if the initial
inclination angle of the disc, $\theta$, and the magnitudes of the disc and
binary angular momentum, $J_{\rm d}$ and $J_{\rm b}$ respectively, satisfy
\begin{equation}
\cos\theta < -\frac{J_{\rm d}}{2J_{\rm b}}.
\label{criterion}
\end{equation}

NKP only considered the zero-frequency (azimuthally symmetric $m=0$) term in
the binary potential. This is a reasonable approach as all other terms induce
oscillatory effects which cancel out on long timescales.

It is natural to assume that coaligned discs are stable, however in the past
counteraligned discs have been incorrectly found to be unstable
\citep{SF1996}. I therefore use a full three dimensional hydrodynamic approach
to check the assumptions in NKP are valid and to confirm that co-- and
counter--alignment are stable. In \sref{cocounter} I discuss the possibility
of both co-- and counter--alignment of the same disc (cf \citealt{LP2006}). In
\sref{sims} I report a simulation of a counteraligning disc and give my
conclusions in \sref{conc}.

\section{Simultaneous co-- {\it and} counter--alignment?}
\label{cocounter}

\citet{LP2006} considered the alignment of a disc and a spinning black
hole. They showed that for a disc where $\theta>\pi/2$ and $J_{\rm d} > 2
J_{\rm h}$, initial counteralignment of the inner disc occurs, followed by
subsequent coalignment of the outer disc. During the alignment of the outer
disc the inner disc retains its retrograde nature and so a warp of significant
amplitude is achieved ($\Delta \theta \sim \pi$). This scenario produces a
disc which is simultaneously counteraligned (in the inner parts) and coaligned
(in the outer parts). For the subsequent evolution of such a disc see
\citet*{Nixonetal2012}.

If however the angular momenta are such that $J_{\rm d} \lesssim J_{\rm h}$
then the disc can only wholly co-- or counter--align and the above process is
impossible. In this section we discuss whether it is feasible to have
circumbinary discs such that both co-- and counter--alignment are
simultaneously possible. This possibility is constrained as the mass of the
disc is limited by self--gravitational collapse and the radius of the disc is
limited by the gravitational sphere of influence of the binary. Therefore
there must be a maximum feasible disc angular momentum.

The disc simultaneously co-- and counter--aligns if (and only if)
$\theta>\pi/2$ and the condition (\ref{criterion}) does not hold. So for
randomly aligned discs, simultaneous co-- and counter--alignment is only
possible if $J_{\rm d} \gtrsim 2 J_{\rm b}$ (assuming $\cos\theta \sim
-1$). Thus I derive the condition for $J_{\rm d} \gtrsim 2 J_{\rm b}$. The
disc angular momentum is
\begin{equation}
  J_{\d} \sim M_{\d}\sqrt{GMR_{\d}}
\label{Jd}
\end{equation}
where $M_{\d}$ is the mass in the disc, $M$ is the total binary mass, $R_{\d}$
is a characteristic radius for the disc and $G$ is the gravitational
constant. We note that the definition of $J_{\rm d}$ in this case is simply
the total angular momentum in the disc. The exact definition of $J_{\rm d}$ is
usually more subtle because the disc takes time to communicate its angular
momentum.

The angular momentum of a binary with eccentricity $e$ is
\begin{equation}
  J_{\b} = \mu\sqrt{GMa\left(1-e^{2}\right)}
\label{Jb}
\end{equation}
where $M$ is the total mass of the binary and $\mu$ is the reduced mass ($\mu
\sim M_2$ for $M_2 \ll M_1$).

So for simultaneous co-- and counter--alignment we require
\begin{equation}
  M_{\d}\sqrt{GMR_{\d}} \gtrsim  2\mu\sqrt{GMa\left(1-e^{2}\right)}.
\label{Jd_JB}
\end{equation}

Self--gravity limits $M_{\d} \lesssim (H/R)M$, so we get
\begin{equation}
  (H/R)M \sqrt{GMR_{\d}} \gtrsim 2\mu\sqrt{GMa\left(1-e^{2}\right)}.
\label{Jd_Jb2}
\end{equation}

After a bit of algebra this tells us that we can only get simultaneous co--
and counter--alignment if
\begin{equation}
  \frac{R_{\d}}{a} \gtrsim 4\left(\frac{R}{H}\right)^{2} \left(\frac{M_2}{M}
  \right)^2\left(1-e^{2}\right)
\label{crit}
\end{equation}

For the most optimistic parameters we have $a\sim 0.1$ pc and the sphere of
influence of the binary $\sim 10$ pc. This gives the LHS as $\lesssim 100$.
For typical disc thickness ($H/R \sim 10^{-3}$) it is clear that binaries with
a low eccentricity ($e \approx 0$) we cannot get simultaneous co-- and
counter--alignment unless the mass ratio is extreme ($M_2/M_1 \lesssim
10^{-2}$). However if the disc is very thick then this is possible. For small
binary mass--ratios the disc mass may be greater than the mass of the
secondary, therefore the hydrodynamic drag on the binary may become
significant \citep{Ivanovetal1999}. If the binary is significantly eccentric,
as predicted for binaries that have accreted through a retrograde disc
\citep{Nixonetal2011a} then simultaneous co-- and counter--alignment may be
possible. I shall return to these possibilities in future work. However these
arguments suggest that it is reasonable to assume an SMBH binary must wholly
co-- or counter--align the disc with the binary plane.

\section{Simulation}
\label{sims}

To confirm the stable counteralignment of a circumbinary disc I perform one
simple simulation. I use the SPH code \textsc{phantom}, a low-memory, highly
efficient SPH code optimised for the study of non-self-gravitating
problems. This code has performed well in related simulations. For example
\citet{LP2010} simulated warped accretion discs and found
excellent agreement with the analytical work of \citet{Ogilvie1999} on the
nature of the internal accretion disc torques.

The implementation of accretion disc $\alpha-$viscosity \citep{SS1973} in
\textsc{phantom} is described in \citet{LP2010}. Specifically, we use the
`artificial viscosity for a disc' described in Sec. 3.2.3 of \citet{LP2010},
similar to earlier SPH accretion disc calculations
\citep[e.g.][]{Murray1996}. The main differences compared to standard SPH
artificial viscosity are that the disc viscosity is applied to both
approaching and receding particles and that no switches are used. The
implementation used here differs slightly from \citet{LP2010} in that the
$\beta^{\rm AV}$ term in the signal velocity is retained in order to prevent
particle interpenetration. The disc viscosity in \textsc{phantom} was
extensively calibrated against the 1D thin $\alpha-$disc evolution in
\citet{LP2010} (c.f. Fig.~4 in that paper) and the disc scale heights employed
here are similar. We use $\alpha^{\rm AV} = 1$ and $\beta^{\rm AV} = 2$ which
at the employed resolution (see below) corresponds to a physical viscosity
$\alpha_{\rm SS} \approx 0.05$. Note that the initial viscosity is slightly
smaller than this, however during the simulation the disc spreads and
particles are accreted and thus the spatial resolution decreases which
increases the viscosity (see eq.~38 of \citealt{LP2010}).

\subsection{Setup}
The simulation has two equal mass sink particles representing the binary (of
total mass unity in code units), on a circular orbit with separation $0.5$ (in
code units). Initially the circumbinary gas disc is flat and composed of $1$
million SPH particles, in hydrostatic equilibrium from $1.0$ to $2.0$ in
radius, and surface density distribution $\Sigma \propto R^{-1}$, set up by
means of the usual Monte--Carlo technique. The vertical hydrostatic
equilibrium corresponds to $H/R = 0.05$ at $R=1$. The equation of state for
the gas is isothermal. The disc is initially tilted at $170^{\circ}$ to the
binary plane. Any gas which falls within a radius of 0.5 from the binary
centre of mass is removed as it no longer has any effect of the alignment of
the circumbinary disc. The disc mass is negligible in comparison to the binary
mass and thus the gravitational back--reaction of the gas on the binary is not
included.

\subsection{Geometry}
The binary is taken to orbit in the $x$--$y$ plane with binary angular
momentum vector in the $z$--direction. We define the `tilt' and the `twist' of
the disc with respect to the binary using Euler angles (e.g. \citealt{BP1975};
\citealt{Pringle1996}), where the unit angular momentum vector in the disc is
described at any radius by
\begin{equation}
  \boldsymbol{\ell} = \left(\cos\gamma \sin\beta, \sin\gamma \sin\beta,
  \cos\beta \right)
\label{geometry}
\end{equation}
with $\beta\left(R,t\right)$ the local angle of disc tilt with respect to the
$z$--axis and $\gamma\left(R,t\right)$ the local angle of disc twist measured
from the $x$--axis.

\subsection{Stable counteralignment}

\begin{figure}
  \begin{flushleft}
    \includegraphics[angle=0,width=0.75\columnwidth]{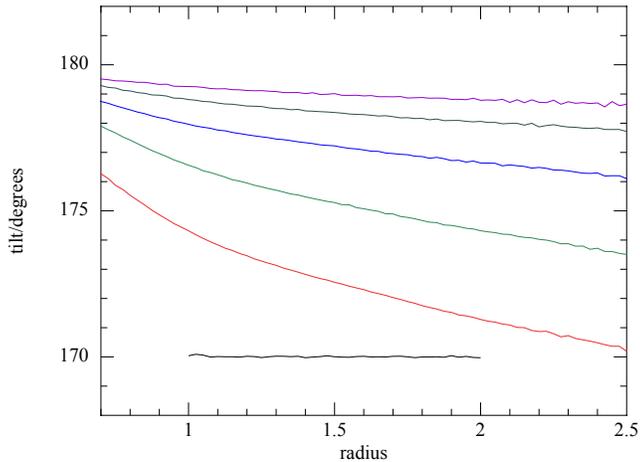}
  \caption{The tilt angle in the disc as a function of radius. This is plotted
    at six different times corresponding to (in units of the dynamical time at
    $R=1$) $t=0$ (black), $t=500$ (red), $t=1000$ (green), $t=1500$ (blue),
    $t=2000$ (grey), $t=2500$ (purple). Initially the disc is flat with a
    global tilt of $170^{\circ}$. In time the disc spreads and
    counteraligns wrt the binary plane. }
  \label{fig:tilt}
  \end{flushleft}
\end{figure}

\begin{figure}
  \begin{flushleft}
    \includegraphics[angle=0,width=0.75\columnwidth]{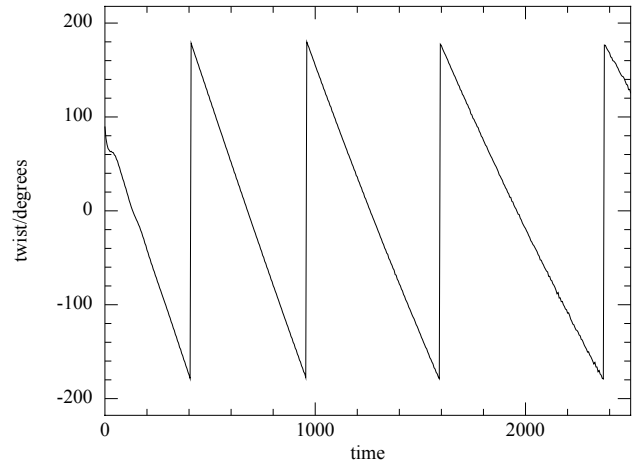}
  \caption{The twist angle (angle from $x$--axis to the line of nodes) at
    $R=1$ in the disc as a function of time. Initially the line of nodes is in
    the $y$--direction and hence the angle is $90^{\circ}$. The binary
    potential induces precession of the gas orbits. The twist is calculated
    between $\pm 180^{\circ}$ which generates the `saw--tooth' structure in
    the plot.}
  \label{fig:twist}
  \end{flushleft}
\end{figure}

As predicted by NKP the dominant effect of the binary on the disc is to induce
precessions in the gas orbits. As the radial range of the disc in this
simulation is only a factor of two, the precession rate changes little between
the inner and outer parts of the disc. However there is still a differential
precession across the disc, which leads to a twist. This causes dissipation
between rings of gas and so a small amplitude warp in the disc
(cf. Fig.~\ref{fig:tilt}). The disc angular momentum vector precesses around
the binary angular momentum vector for the duration of the simulation. In
Fig.~\ref{fig:twist} the twist angle in the disc (at unit radius) is plotted
against time. As the twist angle is defined between $\pm 180^{\circ}$ the plot
has a `saw--tooth' shape, as the disc precesses until the twist is
$-180^{\circ}$ which is then equivalent to $180^{\circ}$.

\begin{figure*}
  \begin{center}
    \subfigure{
       \includegraphics[width=0.3\linewidth]{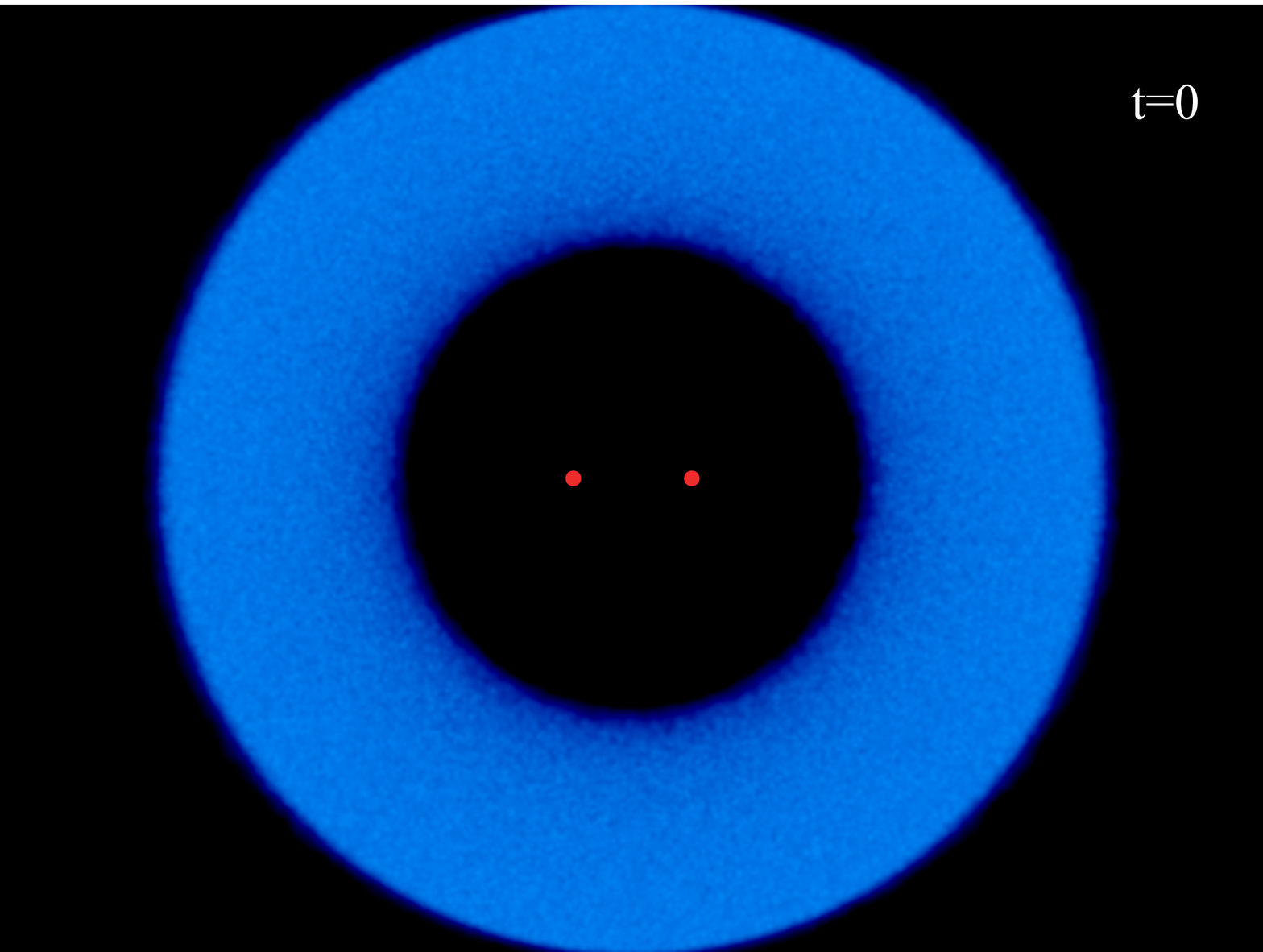}
       \hspace{0.25 cm}
       \includegraphics[width=0.3\linewidth]{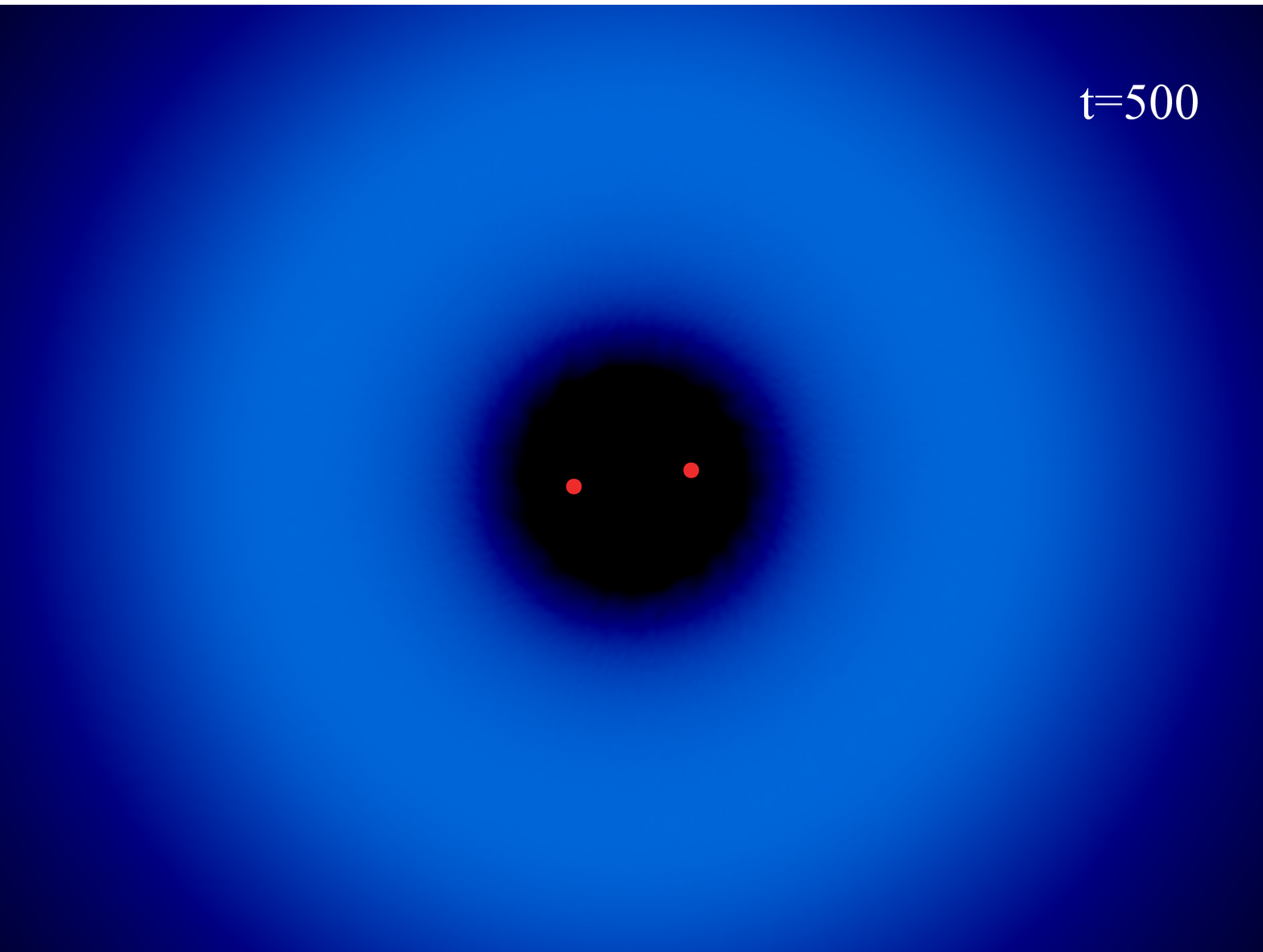} 
       \hspace{0.25 cm}
       \includegraphics[width=0.3\linewidth]{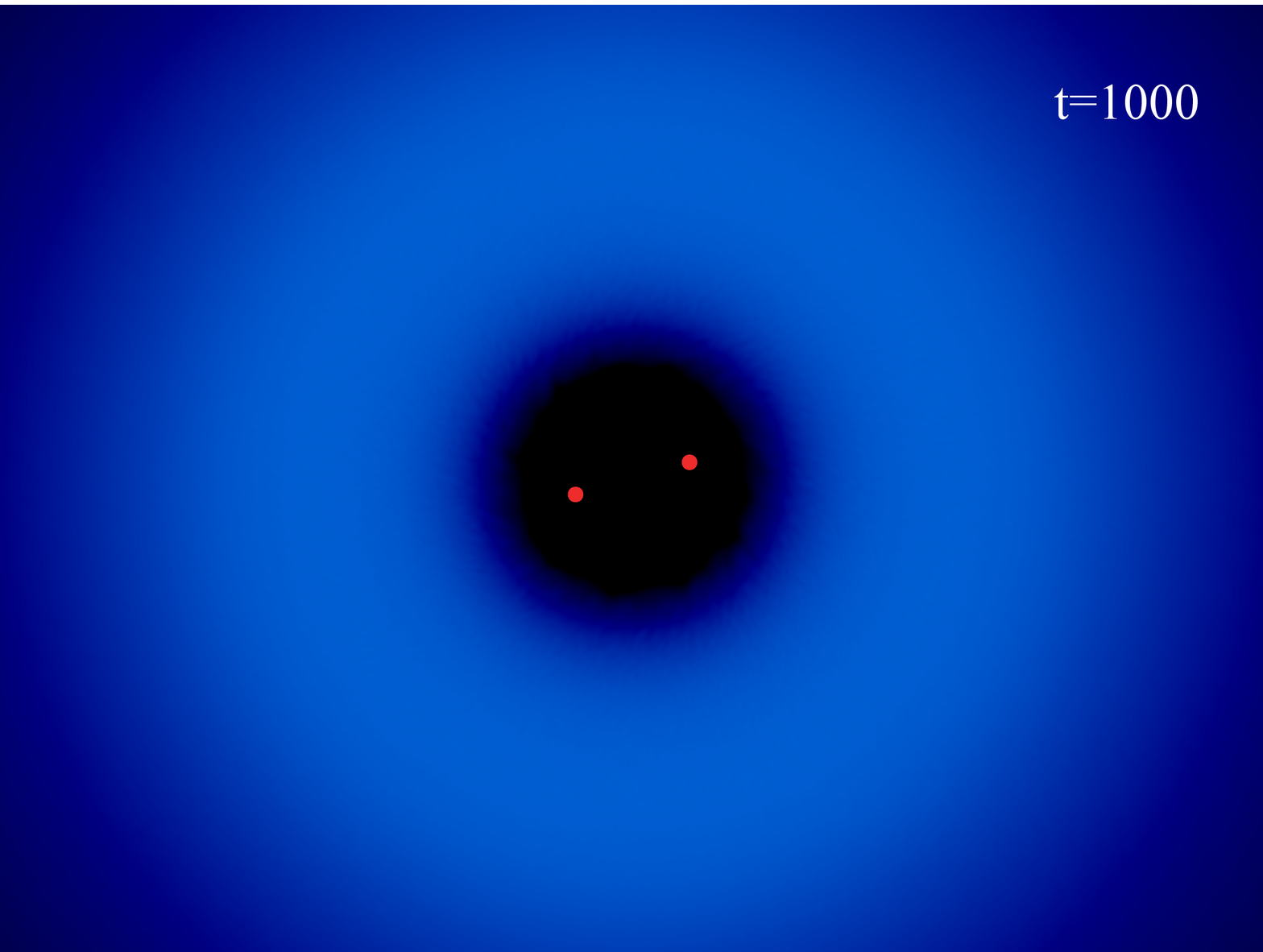}}
    \subfigure{
       \includegraphics[width=0.3\linewidth]{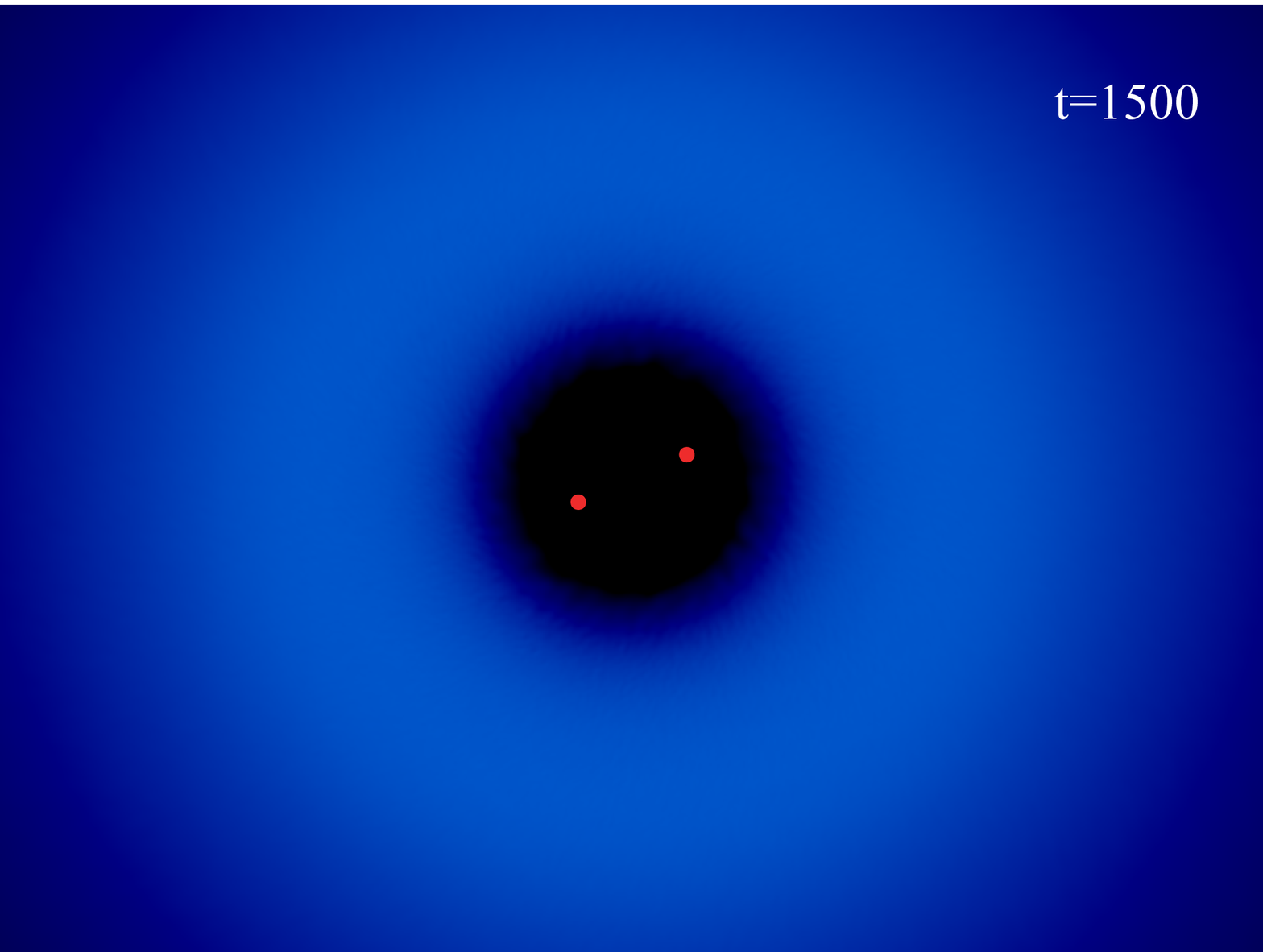}               
       \hspace{0.25 cm}
       \includegraphics[width=0.3\linewidth]{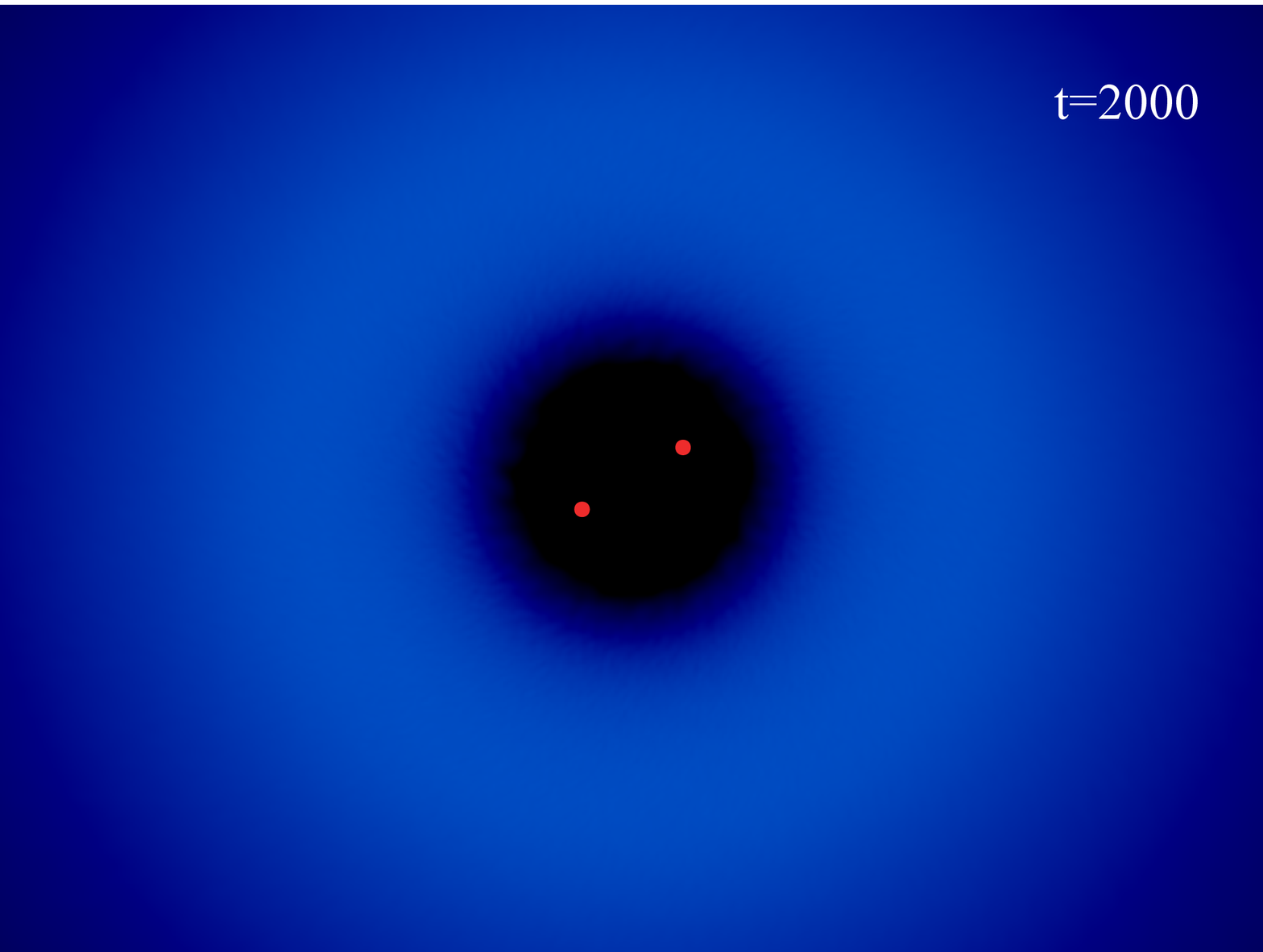} 
       \hspace{0.25 cm}
       \includegraphics[width=0.3\linewidth]{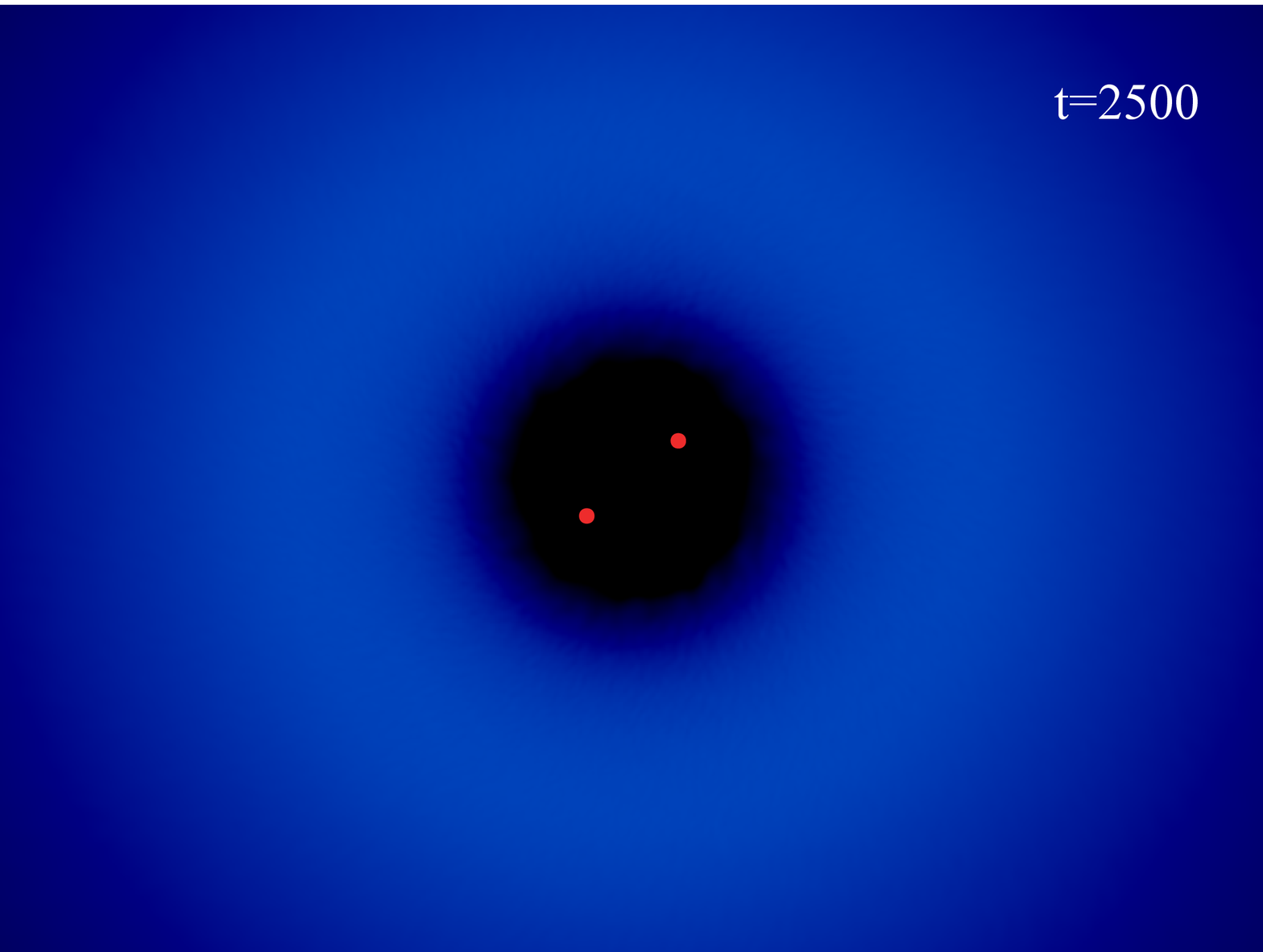}}  
  \end{center}
  \caption{Face--on column density rendering of the disc at various times in
    the simulation. Each component of the binary is represented by a red
    filled circle. Over time the disc spreads and precesses around the
    binary. The precession induces dissipation in the disc which aligns the
    disc with the binary plane.}
  \label{fig:xyrender}
\end{figure*}

\begin{figure*}
  \begin{center}
    \subfigure{
       \includegraphics[width=0.3\linewidth]{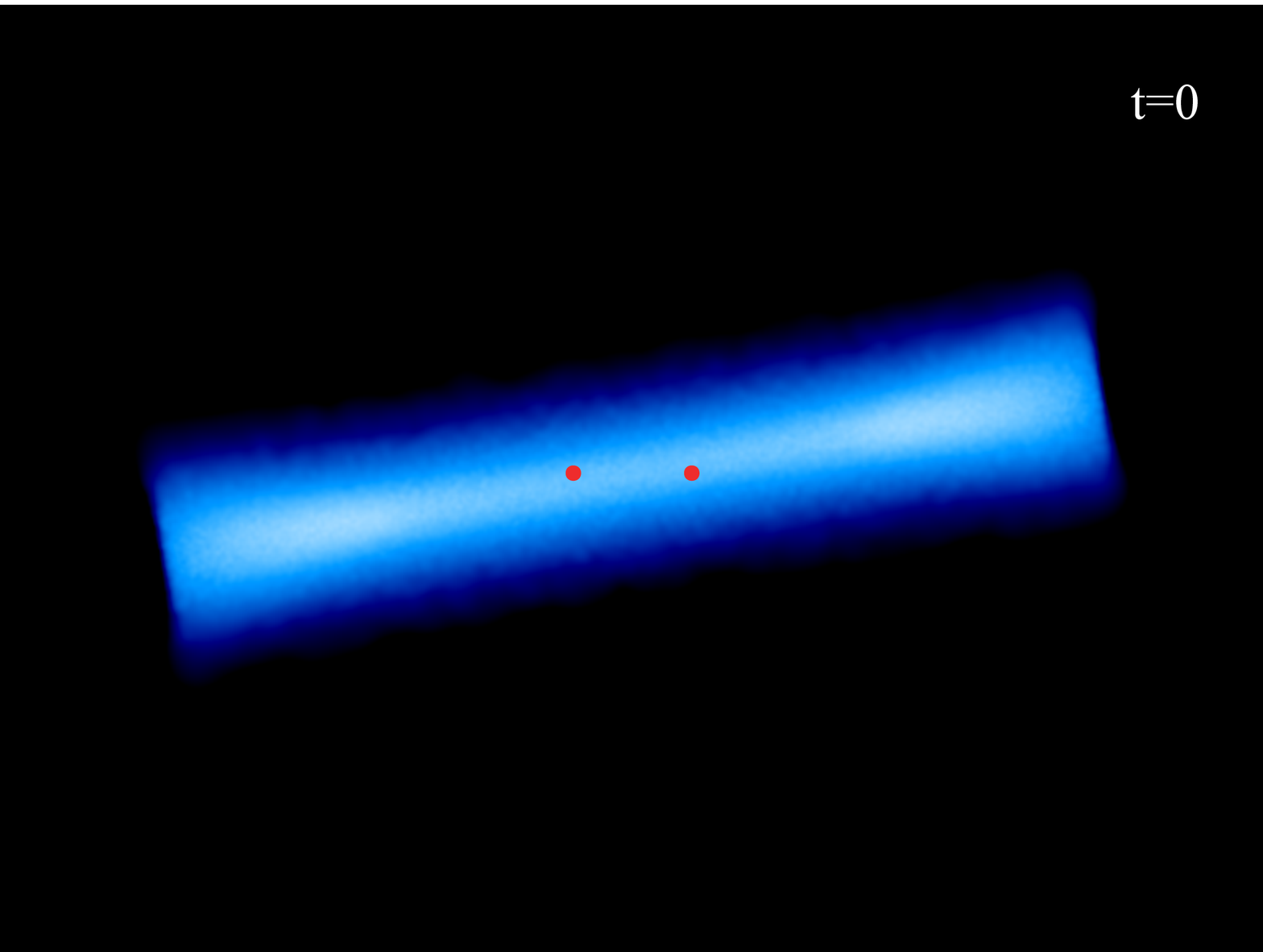}
       \hspace{0.25 cm}
       \includegraphics[width=0.3\linewidth]{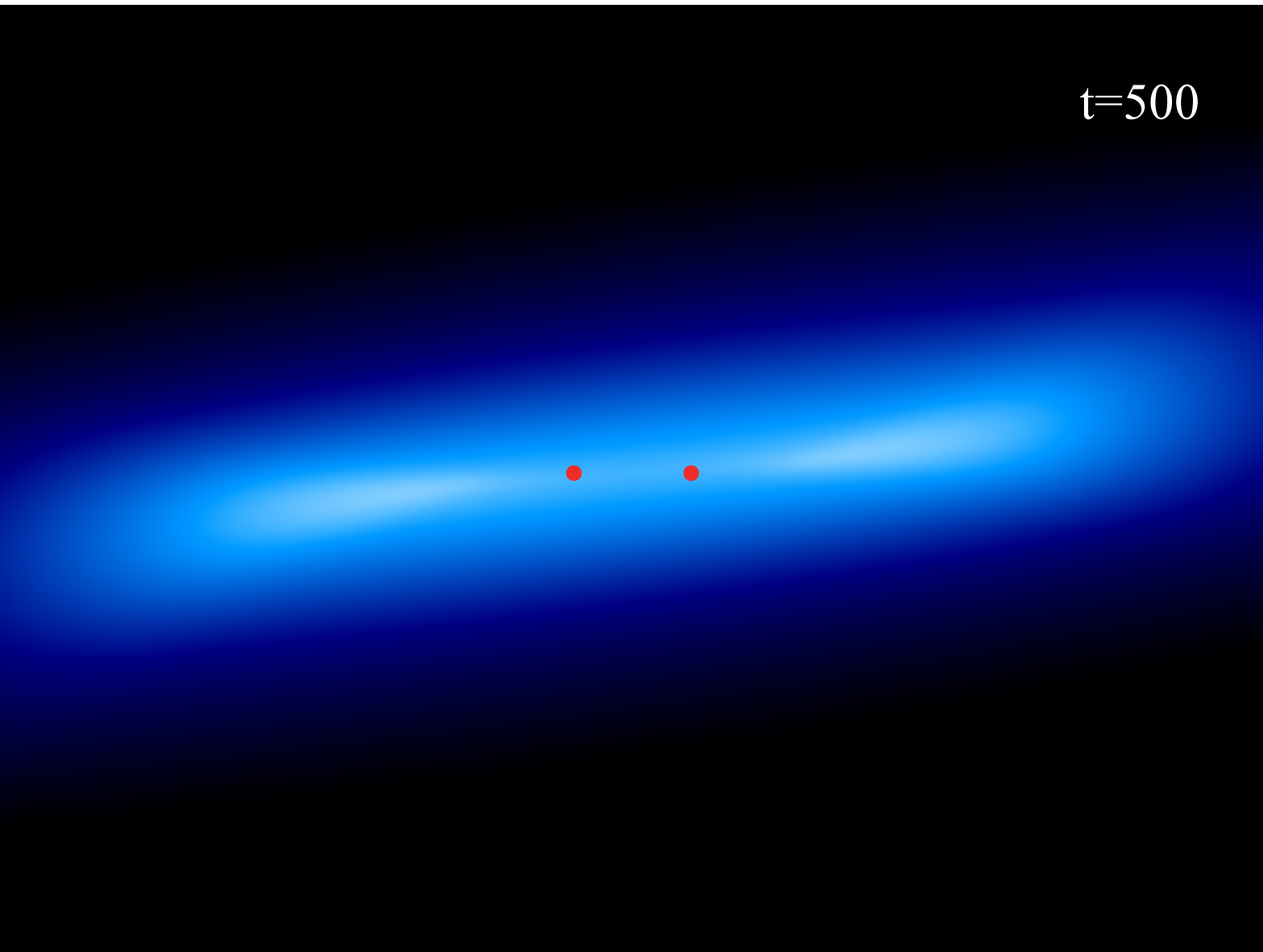} 
       \hspace{0.25 cm}
       \includegraphics[width=0.3\linewidth]{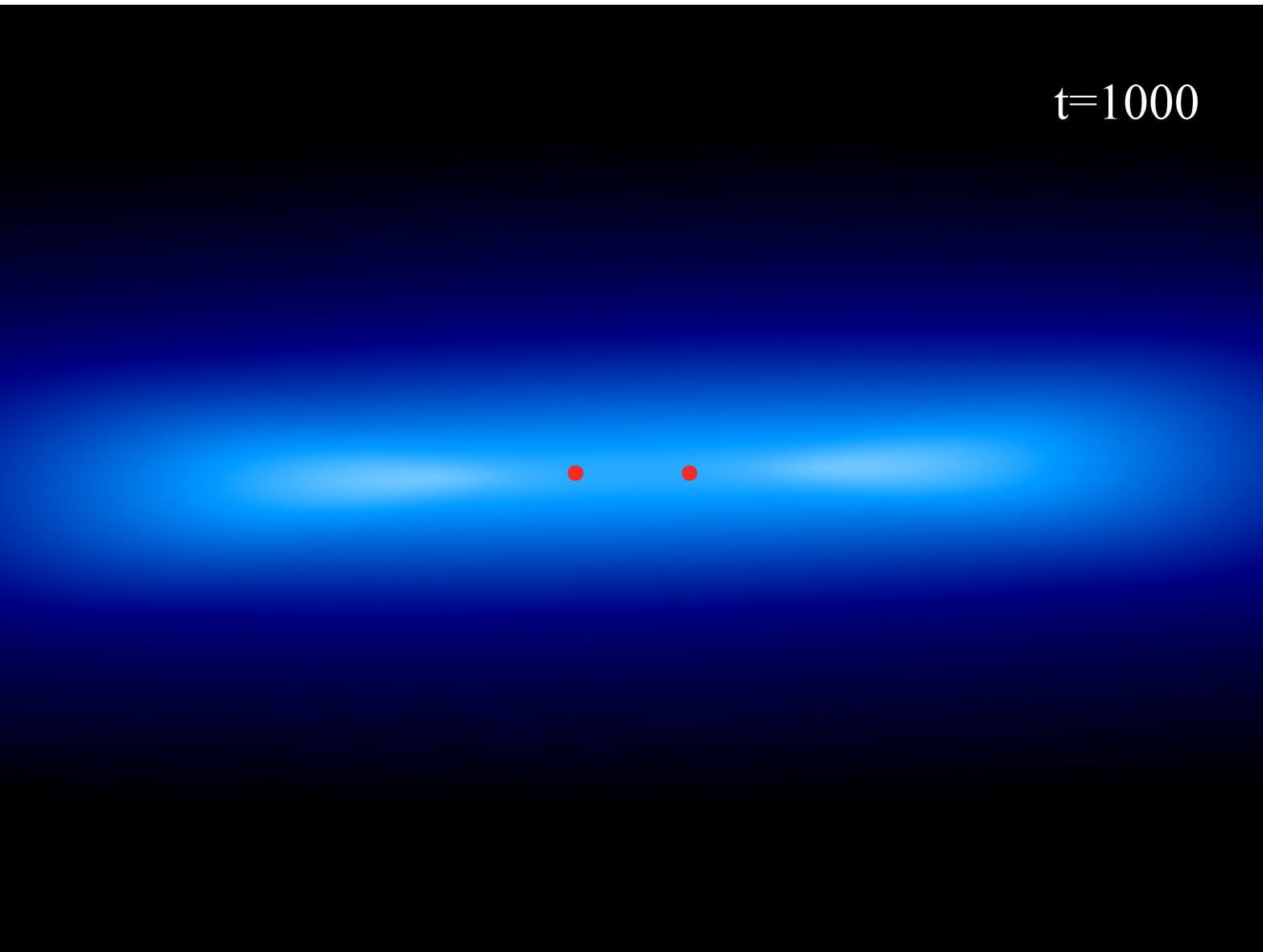}}
    \subfigure{
       \includegraphics[width=0.3\linewidth]{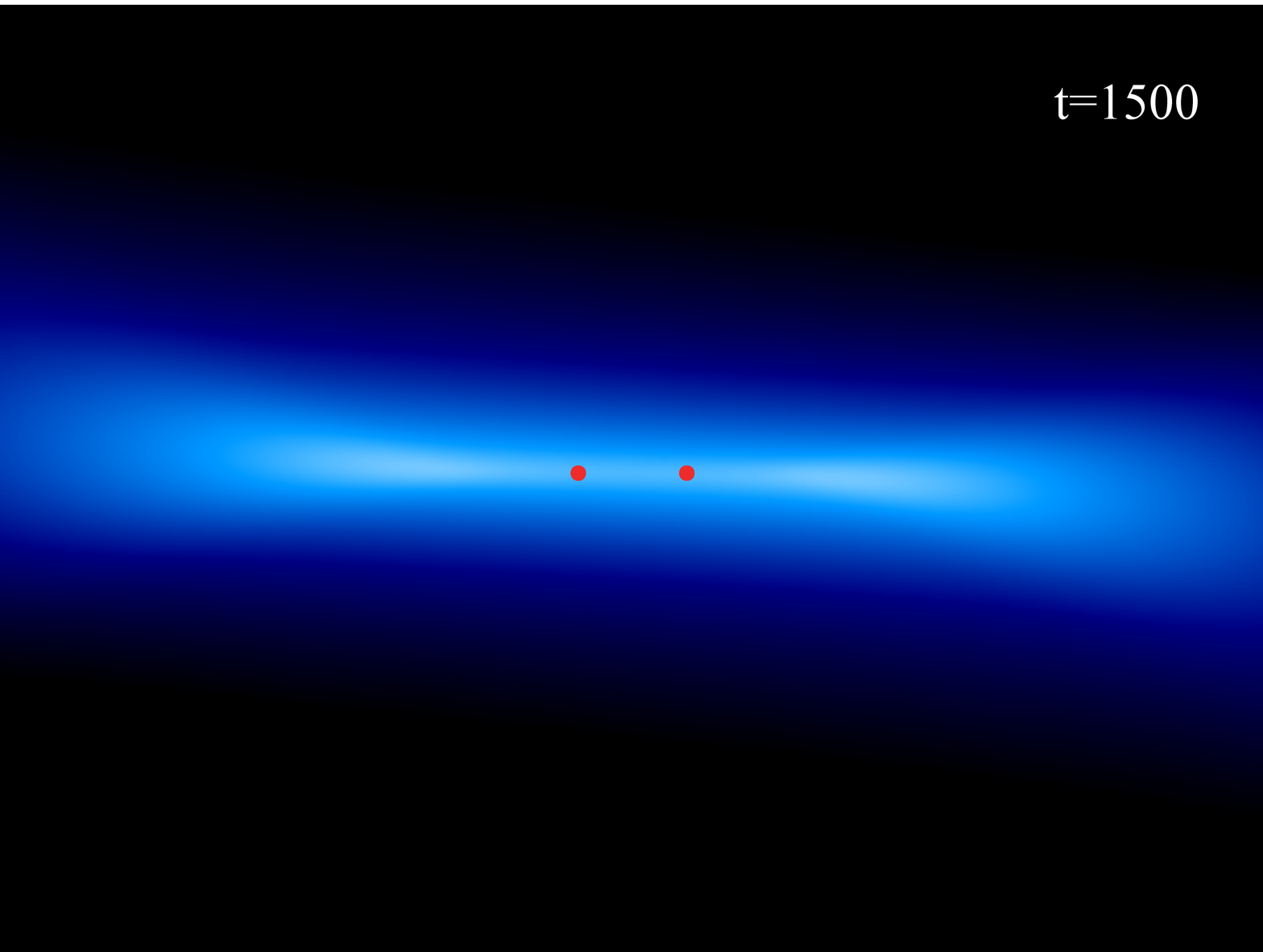}               
       \hspace{0.25 cm}
       \includegraphics[width=0.3\linewidth]{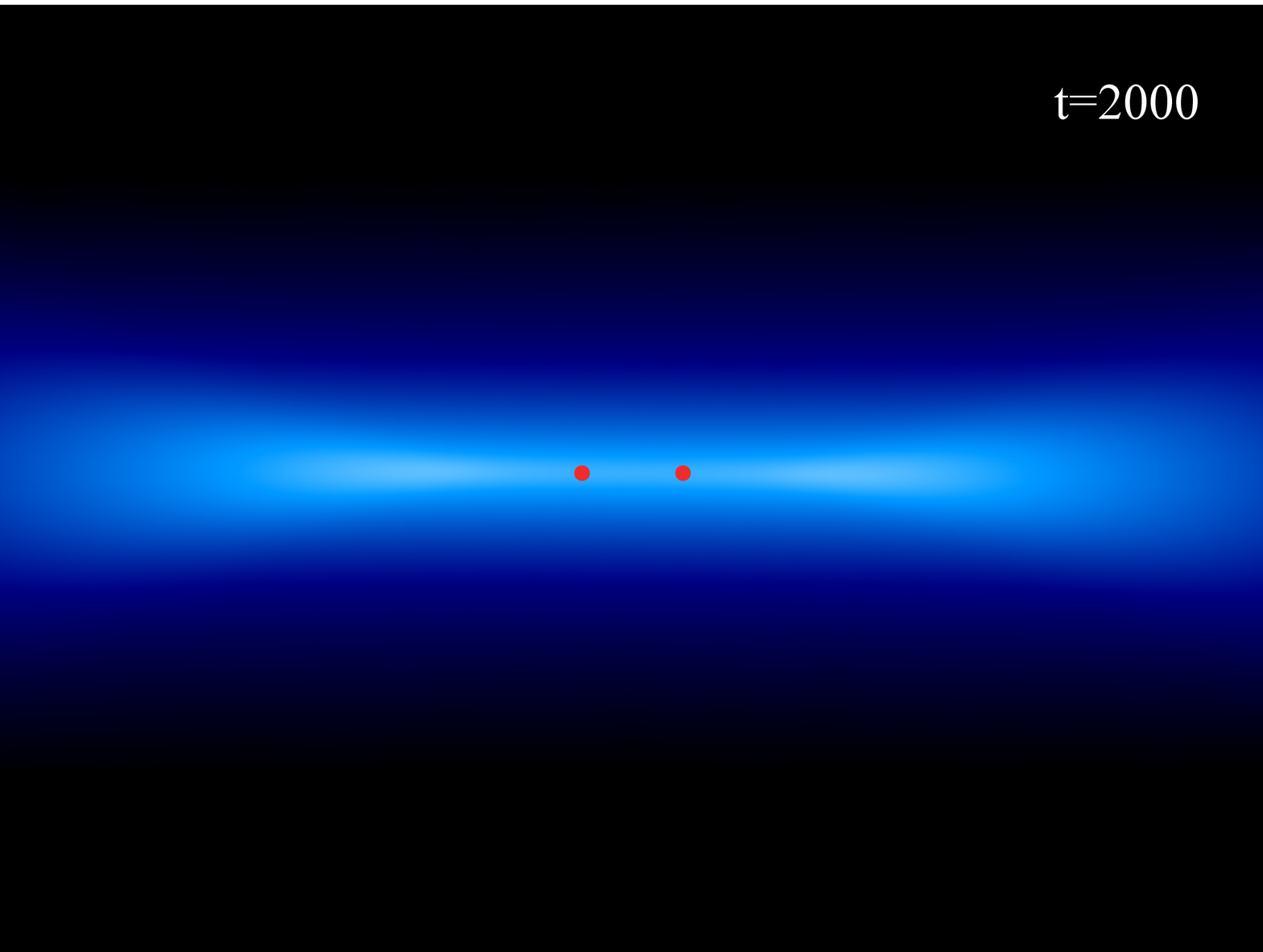} 
       \hspace{0.25 cm}
       \includegraphics[width=0.3\linewidth]{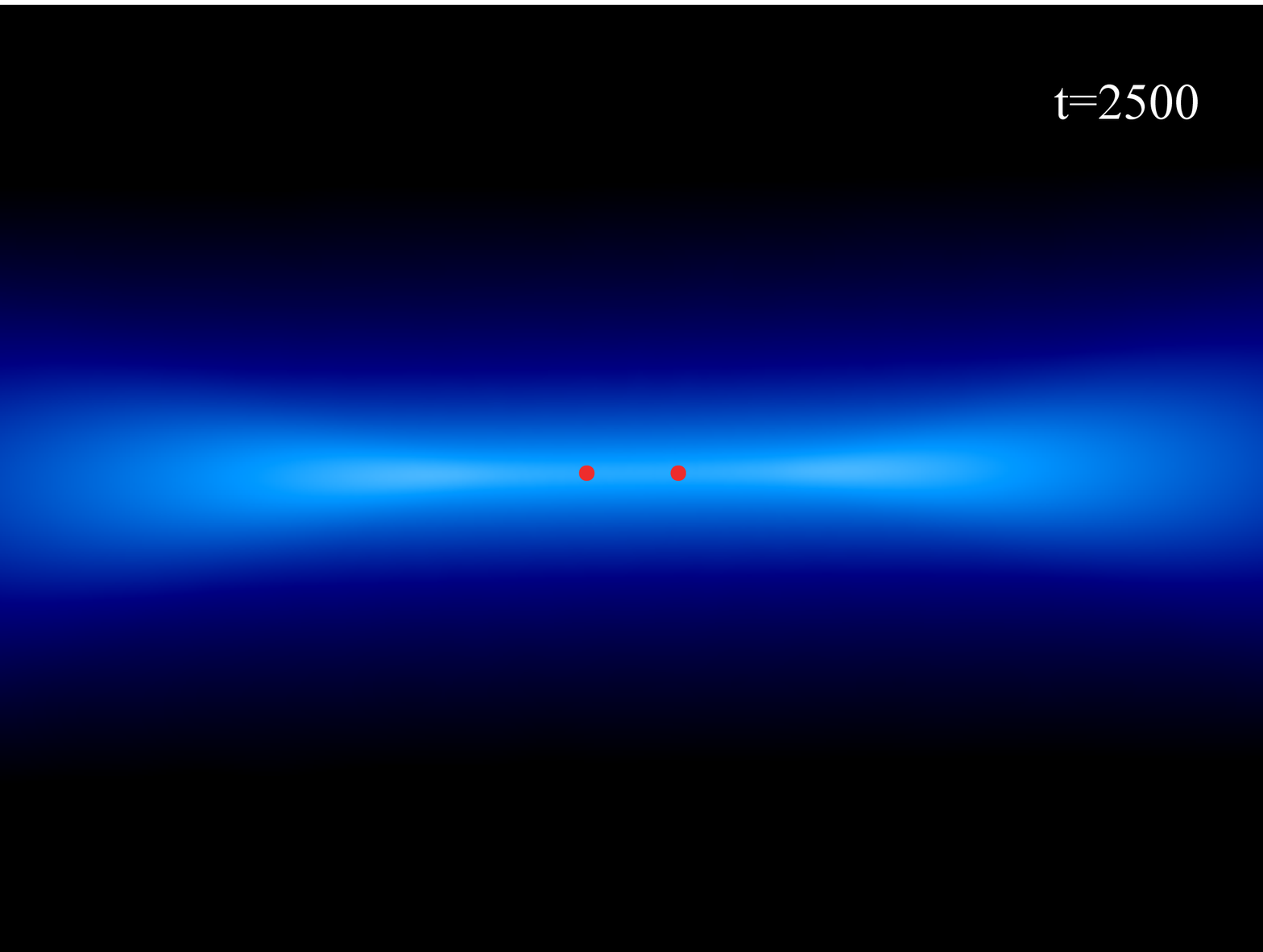}}  
  \end{center}
  \caption{Edge--on column density rendering of the disc at various times in
    the simulation. Each component of the binary is represented by a red
    filled circle. Over time the disc spreads and precesses around the
    binary. The precession induces dissipation in the disc which aligns the
    disc with the binary plane.}
  \label{fig:xzrender}
\end{figure*}

Figs.~\ref{fig:xyrender}~\&~\ref{fig:xzrender} show the disc structures at
various times. Fig.~\ref{fig:xyrender} shows the disc column density viewed
face on, and Fig.~\ref{fig:xzrender} shows it edge on. In
Fig.~\ref{fig:xyrender} we see the usual viscous spreading of the disc, and
also the lack of any resonances in the disc. This is expected in a
counterrotating circumbinary accretion disc (\citealt{PP1977};
\citealt{Nixonetal2011a}). In Fig.~\ref{fig:xzrender} we can see the tilt of
the disc. Initially the disc is tilted by $170^{\circ}$, but then
precesses and counteraligns with the binary plane.

I have also performed simulations (not illustrated) of a moderately eccentric
binary ($e=0.3$) and an unequal mass ratio binary ($M_2/M_1=0.5$) which both display
the same behaviour.

\section{Conclusions}
\label{conc}

In this paper I have shown that for realistic parameters a circumbinary disc
must usually wholly co-- {\it or} counter--align with a binary. However for
extreme mass ratios or high eccentricities the binary may be dominated by the
disc angular momentum (cf. eq.~\ref{crit}). In this case evolution similar to
Fig.~9 of \citet{LP2006} is expected with simultaneous co-- and
counter--alignment of the disc.

I have shown that a disc with an initial
inclination of $170^{\circ}$ to the binary plane stably
counteraligns. Circumbinary discs, with $J_{\rm d}
\ll J_{\rm b}$ and an initial misalignment angle of $> 90^{\circ}$
counteralign with respect to the binary. If the disc angular momentum is not
negligible, \citet{Kingetal2005} showed that the condition for
counteralignment is
\begin{equation}
\cos\theta < -\frac{J_{\rm d}}{2J_{\rm b}}.
\end{equation}

A counteraligned circumbinary disc is efficient at shrinking the binary as it
directly absorbs negative angular momentum when capturing gas into
circumprimary or circumsecondary discs \citep{Nixonetal2011a}. This
interaction increases the binary eccentricity. \citet{Nixonetal2011a} show
that the timescale to increase the eccentricity from zero to unity is $\sim
M_2/{\dot M}$ where ${\dot M}$ is the mass inflow rate through the retrograde
circumbinary disc. Once the eccentricity is high enough gravitational wave
losses will drive the binary to coalescence.

\section*{Acknowledgments}
\label{acknowledgments}
I thank Andrew King and Daniel Price for providing useful feedback on the
manuscript. I also thank the referee for useful suggestions. CJN holds an STFC
postgraduate studentship. Research in theoretical astrophysics at Leicester is
supported by an STFC Rolling Grant. I acknowledge the use of \textsc{splash}
\citep{Price2007} for the rendering of the figures. This research used the
ALICE High Performance Computing Facility at the University of Leicester.
Some resources on ALICE form part of the DiRAC Facility jointly funded by STFC
and the Large Facilities Capital Fund of BIS.

\bibliographystyle{mn2e} 
\bibliography{nixon}

\end{document}